1  Title: Mandated data archiving greatly improves access to research data

3  Authors: Timothy H. Vines[1,2,*], Rose L. Andrew[1], Dan G. Bock[1], Michelle T. Franklin[1,3],

4  Kimberly J. Gilbert[1], Nolan C. Kane[1,4], Jean-Sébastien Moore[1], Brook T. Moyers[1], Sébastien

5  Renaut[1], Diana J. Rennison[1], Thor Veen[1], Sam Yeaman[1]

7  Affiliations: [1]Biodiversity Department, University of British Columbia, 6270 University Blvd

8  Vancouver BC, Canada, V6T 1Z4. [2]Molecular Ecology Editorial Office, 6270 University Blvd

9  Vancouver BC, Canada, V6T 1Z4. [3]Department of Biological Sciences, Simon Fraser

10  University, 8888 University Drive, Burnaby, British Columbia, Canada, V5A 1S6. [4]Ecology and

11  Evolutionary Biology Department, University of Colorado at Boulder, Boulder, CO, USA, 80309

13  [*]Author for correspondence: vines@zoology.ubc.ca. Tel: 1 778 989 8755 Fax: 1 604 822 8982

15  Short title: Archiving policies and access to research data

16

17

18  Non standard abbreviations:

19  PLoS – Public Library of Science

20  BMC – BioMed Central

21  BJLS – Biological Journal of the Linnean Society

22  TAG – Theoretical and Applied Genetics

23  IF – Impact Factor

24

25  Abstract: The data underlying scientific papers should be accessible to researchers both now and
26  in the future, but how best can we ensure that these data are available? Here we examine the
27  effectiveness of four approaches to data archiving: no stated archiving policy, recommending
28  (but not requiring) archiving, and two versions of mandating data deposition at acceptance. We
29  control for differences between data types by trying to obtain data from papers that use a single,
30  widespread population genetic analysis, STRUCTURE. At one extreme, we found that mandated
31  data archiving policies that require the inclusion of a data availability statement in the manuscript
32  improve the odds of finding the data online almost a thousand-fold compared to having no
33  policy. However, archiving rates at journals with less stringent policies were only very slightly
34  higher than those with no policy at all. We also assessed the effectiveness of asking for data
35  directly from authors and obtained over half of the requested datasets, albeit with about 8 days'
36  delay and some disagreement with authors. Given the long-term benefits of data accessibility to
37  the academic community, we believe that journal-based mandatory data archiving policies and
38  mandatory data availability statements should be more widely adopted.

39
40

43
44

45  Archiving the data underlying scientific papers is an essential component of scientific

46  publication and its subsequent reproducibility [*1-3*], but very few papers actually make the

47  underlying data available [*4*]. In response to this gap between the needs of science and author

48  behavior, a number of journals have introduced data archiving policies. Here, we evaluate the

49  effectiveness of these policies by comparing journals that have no stated data archiving policy,

50  journals that recommend data archiving, and journals that mandate archiving prior to publication.

51  Journals that mandate data archiving fall into two further subgroups: those that require an

52  explicit data availability statement and those that do not. We ask two questions: (1) does having

53  any kind of data archiving policy improve the likelihood of the data being available online, and

54  (2) does the type of data archiving policy have any effect the likelihood of obtaining the data?

55

56  We recently assembled datasets from a range of journals for a study of the reproducibility of

57  commonly used population genetic analyses [*5*]. Here, we use this opportunity to examine

58  whether data archiving policy (or lack thereof) was associated with the proportion of datasets we

59  were able to obtain from a journal. As papers within even a single journal contain many different

60  types of data, we restricted both this and our reproducibility study to articles using the population

61  genetics program STRUCTURE [*6*]. We chose STRUCTURE because it is widely used in ecology and

62  evolution, and because the underlying data is a table of microsatellite, Amplified Fragment

63  Length Polymorphism or Single Nucleotide Polymorphism genotypes, and for the ease of

64  archiving this type of dataset online. For example, the data could be uploaded as supplemental

65  material, or archived on the Dryad repository [*7*]. Dryad was established in 2010 for the

66  preservation of a wide range of data types associated with ecology or evolution articles, and is

67  often used to archive STRUCTURE datasets.

68

69  *Data collection*

70

71  We used Web of Science to identify papers published in 2011 or 2012 that cited the original

72  description of STRUCTURE [*6*]. We selected journals for each of the four journal categories

73  described above, and excluded those that had less than five eligible papers. We complemented

74  our list of papers by searching for additional articles that used STRUCTURE on the journal website.

75  Papers that used DNA sequence data were excluded, as preparing raw sequence data from e.g.

76  GenBank for re-analysis with STRUCTURE was found to be very time consuming.

77

78  We found four eligible journals with no stated data archiving policy: *Conservation Genetics*,

79  *Crop Science*, *Genetica*, and *Theoretical and Applied Genetics* (*TAG*).

80

81  There were four eligible journals that had some sort of data archiving policy but stopped short of

82  mandating archiving for all data (*BMC Evolutionary Biology*, *Biological Journal of the Linnean*

83  *Society* (*BJLS*), *Journal of Heredity* and *PLoS One*). These policies were retrieved from the

84  author guidelines in mid 2011 and are available on Dryad (doi:10.5061/dryad.6bs31). The latter

85  three journals ask that the data be placed onto an online archive whenever one exists. For

86  STRUCTURE data, Dryad is the most commonly used repository, and indeed the policies of the last

87  two journals (*J. Heredity* and *PLoS One*) explicitly mention Dryad. There is thus an expectation

88  for three of these four journals the data should be available somewhere online, most likely on

89  Dryad. For *BMC Evolutionary Biology* the data will only be online if the authors have decided to

90  share it at publication. The individual policies are as follows:

91

92  First, *BMC Evolutionary Biology* states that "submission … implies that readily reproducible

93  materials described in the manuscript, including all relevant raw data, will be freely available to

94  any scientist wishing to use them for non-commercial purposes", and at that time did not require

95  that data appear in an online archive. This policy has been in place since 2009.

96

97  Second, the *Biological Journal of the Linnean Society* has the policy "we recommend that data

98  for which public repositories are widely used, and are accessible to all, should be deposited in

99  such a repository prior to publication." This policy was introduced in January 2011 and we hence

100  only considered papers submitted after this date.

101

102  Third, *J. Heredity* "endorses the principles of the Joint Data Archiving Policy [see below] in

103  encouraging all authors to archive primary datasets in an appropriate public archive, such as

104  Dryad, TreeBASE, or the Knowledge Network for Biocomplexity". As with *BJLS*, this policy

105  was introduced in January 2011 and we hence only considered papers submitted after this date.

106

107 Fourth, *PLoS One* has had a policy on data sharing in place since 2008, and one statement is as

108 follows: "If an appropriate repository does not exist, data should be provided in an open access

109 institutional repository, a general data repository such as Dryad, or as Supporting Information

110 files with the published paper."

111

112 Finally, there were four journals that adopted a mandatory data archiving policy (known as the

113 Joint Data Archiving Policy or JDAP [*1*]), which states "[Journal X] requires, as a condition for

114 publication, that data supporting the results in the paper should be archived in an appropriate

115 public archive". For these journals we excluded papers submitted before the policy came into

116 force: January 2011 for *Molecular Ecology*, *Journal of Evolutionary Biology*, and *Evolution*, and

117 March 2011 for *Heredity*. Of these four, two (*Molecular Ecology* and *Heredity*) additionally

118 require that authors include a data availability statement within each accepted manuscript; these

119 sections describe the location (typically the database and accession numbers) of all publicly

120 available data.

121

122 For all 229 eligible papers we then checked whether the STRUCTURE genotype data was available

123 either as supplemental material or elsewhere online, such as on the Dryad archive [*7*]. Our results

124 are shown in Table 1 and Figure 1, and the data and R code used in the analysis are archived at

125 doi:10.5061/dryad.6bs31.

126

127 *Statistical analysis*

128

129 To evaluate the statistical support for an association between the presence/absence of an

130 archiving policy and whether or not the STRUCTURE data could be found online, we fitted a

131 mixed effects logistic regression. The response variable was whether or not the data from a paper

132 was available online, coded as '0' for not available and '1' for available. The predictor variable

133 was either 'no policy' or 'archiving policy', and journals were included as a random effect within

134 each category.

135

136  Having any sort of archiving policy did lead to a significant improvement in the probability of
137  the data being online (likelihood ratio test statistic = 4.27, p= 0.038), such that the odds of
138  getting the data were about 25 times higher (95% confidence interval: 1.5 to 416.7).
139
140  We then tested how well each type of archiving policy compared to having no policy at all. As
141  above, we used a mixed effects logistic regression. Again, the response variable was whether or
142  not the data from a paper was available online, coded as '0' for not available and '1' for
143  available. The predictor variable was policy type, and the categories were 'no policy',
144  'recommend archiving', 'mandate archiving, no data statement' and 'mandate archiving, with
145  data statement'. Journals were a random effect within each policy type. The overall model found
146  that policy type did have a very significant effect on data availability (likelihood ratio test
147  statistic = 28.06, p<0.001).
148
149  Since this is a logistic model, we can readily calculate the effect that the different policy types
150  have on the likelihood that the data will be available. We explore these odds for each type of
151  policy below, using 'no policy' as the baseline.
152
153  Having a 'recommend archiving' policy made it 3.6 times more likely that the data were online
154  compared to having no policy. However, the 95% confidence interval overlapped with 1 (0.96 to
155  13.6), and hence this increase in the odds is not significant. Overall, recommending data
156  archiving is only marginally more effective than having no policy at all.
157
158  The data was 17 times more likely to be available online for journals that had adopted a
159  mandatory data archiving policy but did not require a data accessibility statement in the
160  manuscript. This odds ratio was significantly greater than 1 (95% confidence interval: 3.7 to
161  79.6).
162
163  For 'mandate archiving' journals where a data accessibility statement is required in the
164  manuscript, the odds of finding the data online were 974 times higher compared to having no
165  policy. The 95% confidence interval on these odds is very wide (97.9 to 9698.8), but nonetheless

166  shows that the combination of a mandatory policy and an accessibility statement is much more
167  effective than any other policy type.
168
169
170  *Requesting data directly from authors*
171
172  A number of the 'recommend archiving' policies state that the data should also be freely
173  available from the authors by request (see the 'Journal Policies' file at doi:10.5061/dryad.6bs31),
174  and hence we wanted to evaluate whether obtaining data directly from authors is an effective
175  approach. Part of the dataset collection for our reproducibility study [*5*] involved emailing
176  authors of papers from two of the 'recommend archiving' journals (*BMC Evolutionary Biology*
177  and *PLoS One*) and requesting their STRUCTURE input files. Here, we examine how often these
178  requests led to us obtaining the data. We did not email the authors of articles where the data were
179  already available online. A detailed description of our data request process appears on Dryad
180  (doi:10.5061/dryad.6bs31), but we essentially contacted corresponding and senior authors of
181  each paper up to three times over a three week period, and recorded if and when the data were
182  received.
183
184  We obtained data directly from the authors for seven of the 12 eligible papers in *BMC*
185  *Evolutionary Biology*, and 27 datasets from 45 papers from *PLoS One* (Table 1). All seven of the
186  *BMC Evolutionary Biology* datasets arrived between eight and 14 days after our initial request.
187  Ten of the *PLoS One* datasets came within a week, 13 came between eight and 14 days, and four
188  arrived between 15 and 21 days. Unlike the online data, which could generally be obtained
189  within a few minutes, the requested datasets took a mean of 7.7 days to arrive, with one author
190  responding that the dataset had been lost in the year since publication. More than one email had
191  to be sent to the corresponding and/or senior author for 53% of papers, and the authors of 29% of
192  the papers did not respond to any of our requests. No data were received more than 21 days after
193  our initial request. We also note that requesting data via email did upset some authors,
194  particularly when they were reminded of the journal's data archiving policy or when multiple
195  emails were sent.
196

197 Our average return of 59% in an average of 7.7 days is markedly better than has been reported in
198 similar studies: Wicherts *et al.* [*8*] received only 26% of requested datasets after six months of
199 effort with authors of 141 psychology articles, and Savage and Vickers [*9*] received only one of
200 ten datasets requested from papers in *PLoS Medicine* and *PLoS Clinical Trials*. In a 1999 study,
201 Leberg and Neigel [*10*] emailed the authors of 30 papers that contained an incomplete
202 description of their sequence dataset, but received the requested data from just one of them.
203 Since the latter study and ours both involved the evolutionary biology community, it appears that
204 attitudes to data sharing have improved dramatically over the last decade. However, the two
205 more recent studies that used human data still had low success rates, perhaps because privacy
206 and consent issues are a significant impediment to data sharing in these fields.
207
208
209 *Conclusions*
210
211 Our results demonstrate that journal-based data archiving policies can be very effective in
212 ensuring that research data are available to the scientific community, especially when journals
213 require that a data accessibility statement appear in the manuscript. The 'recommend archiving'
214 group of journals encompassed the broadest spread of policy types, yet as a whole only had 10 of
215 89 datasets available. The policies range from a simple "Submission … implies that … all
216 relevant raw data, will be freely available to any scientist wishing to use them for non-
217 commercial purposes" at *BMC Evolutionary Biology* to an endorsement of the full Joint Data
218 Archiving Policy at *J. Heredity*. However, none of these policies led to more than 23% of the
219 data being available online (at *BJLS*), and there was no significant difference between the
220 success of this policy type and having no policy at all.
221
222 Interestingly, *PLoS One*'s very comprehensive policy, which is over 1000 words long and
223 contains statements like "data should be provided in an open access institutional repository, a
224 general data repository such as Dryad, or as Supporting Information files with the published
225 paper" was only marginally more effective than *BMC Evolutionary Biology*'s simple request that
226 the data be freely available, with 11% and 7% of the data online, respectively.
227

228     The difference between *PLoS One* and the 'mandate archiving' journals may arise because the

229     wide breadth of subject areas in *PLoS One* precludes having a policy with the bald simplicity of

230     the Joint Data Archiving Policy: "[Journal X] requires, as a condition for publication, that data

231     supporting the results in the paper should be archived in an appropriate public archive". Even

232     though the portion of *PLoS One*'s author community that uses STRUCTURE broadly overlaps with

233     the authors of the papers in the JDAP journals, it may be that the lack of a single strong

234     statement leads to much lower compliance. One simple remedy for this situation might be the

235     introduction of a mandatory data accessibility statement in all manuscripts.  For fields where

236     archiving is not (yet) standard practice, this could state that the data were available from the

237     authors, but in fields where archiving is expected the authors would indicate where their data

238     were available online.

239

240     More broadly, a study by Piwowar and Chapman [*11*] on 397 microarray datasets from 20

241     journals also found that having a 'strong' (i.e. close to mandatory) data archiving policy led to a

242     high proportion (>50%) of the datasets being available online. Journals that had a 'weak' policy

243     (i.e. recommended archiving) had just over 30% of microarray datasets available, and journals

244     with no policy had only about 20% availability. Furthermore, they also found that a journal with

245     an Impact Factor (IF) of 15 was 4.5 times more likely to have the microarray data online than a

246     journal with an IF of 5. We find a similar effect in our data: using the 2010 Impact Factors, we

247     were 3.2 times more likely to find the data online for a journal with an IF of 5.0 (the average IF

248     of the JDAP journals) compared to those with an IF of 2.2 (the average IF of the 'no policy'

249     journals); details of this analysis are available at doi:10.5061/dryad.6bs31. We are able to

250     exclude higher Impact Factor as the primary cause of the high rate of data archiving in the JDAP

251     journals: in 2010 (before the mandatory archiving policy was introduced), none of the 27 eligible

252     papers in the *Journal of Evolutionary Biology*, *Evolution* or *Heredity* had archived their data,

253     even though their Impact Factors were essentially the same in 2010 and 2011 (*Molecular*

254     *Ecology* recommended archiving in 2010 and was excluded from this comparison). This result

255     suggests that the introduction of the JDAP policy in 2011 was primarily responsible for the

256     abrupt rise in the proportion of articles in these three journals that archived their data. However,

257     it is possible that Impact Factor still plays a role, as only journals with a high IF may feel able to

258 introduce stringent archiving policies. The positive effects of a strongly worded data archiving
259 statement were also confirmed by a much larger study involving 11603 microarray datasets [*12*].
260

261 Requesting data directly from authors can also provide access to research data, but this approach
262 can be hampered by delays and the potential for disagreement between requester and the authors.
263 Furthermore, the availability of datasets directly from authors will only decrease as time since
264 publication increases. This is particularly true when researchers leave science or when data that
265 are stored on lab computers or websites get misplaced [*13, 14*].
266

267 Even though our results strongly emphasize the value of public databases for archiving scientific
268 data, these databases do require ongoing financial support; this money may come from funding
269 agencies, journal publishers, libraries or even individual researchers. A recent study put the cost
270 of running the Dryad database at around $400,000 per annum; these costs include the
271 maintenance of their archive and the addition and curation of an extra 10,000 datasets per year.
272 For comparison, the same amount spent by a funding agency on basic research would generate
273 about 16 new publications [*15*]. Given that the long-term availability of these data allows for
274 meta-analyses, the checking of previous results, and not collecting the same data again, money
275 spent on data archiving is extremely cost effective. In light of all these advantages, we believe
276 that journal-based mandatory data archiving policies and data accessibility statements should be
277 more widely adopted.
278

315  Acknowledgments: We thank Heather Piwowar, Loren Rieseberg, Phil Davis and Mike Whitlock
316  for comments on an earlier version of the manuscript, and Arianne Albert for help with the
317  statistics. We would also like to express our gratitude to the many authors who shared their data
318  with us.
319

320     Table 1. The number of eligible articles per journal, and the number for which data were obtained from online databases.

321

| Policy | Journal | No. eligible papers | No. data online |
|---|---|---|---|
| No Policy | *Conservation Genetics* | 47 | 1 |
| | *Crop Science* | 12 | 1 |
| | *Genetica* | 8 | 1 |
| | *T.A.G.* | 21 | 0 |
| Recommend data archiving | *BMC Evolutionary Biol.* | 13 | 1 |
| | *B.J.L.S.* | 13 | 3 |
| | *J. Heredity* | 12 | 0 |
| | *PLoS One* | 51 | 6 |
| Mandatory data archiving | *J. Evolutionary Biology* | 10 | 3 |
| | *Evolution* | 6 | 3 |
| | *Heredity* | 7 | 7 |
| | *Molecular Ecology* | 28 | 27 |

322

323

324 Figure 1. Percentage of eligible papers published in 2011 that made their data available online, by journal. The number of eligible papers is
325 shown above each column. Within the 'mandate archiving' group, 'data statement' denotes the journals that require a data accessibility
326 statement in the manuscript, and 'no data statement' denotes those that do not.

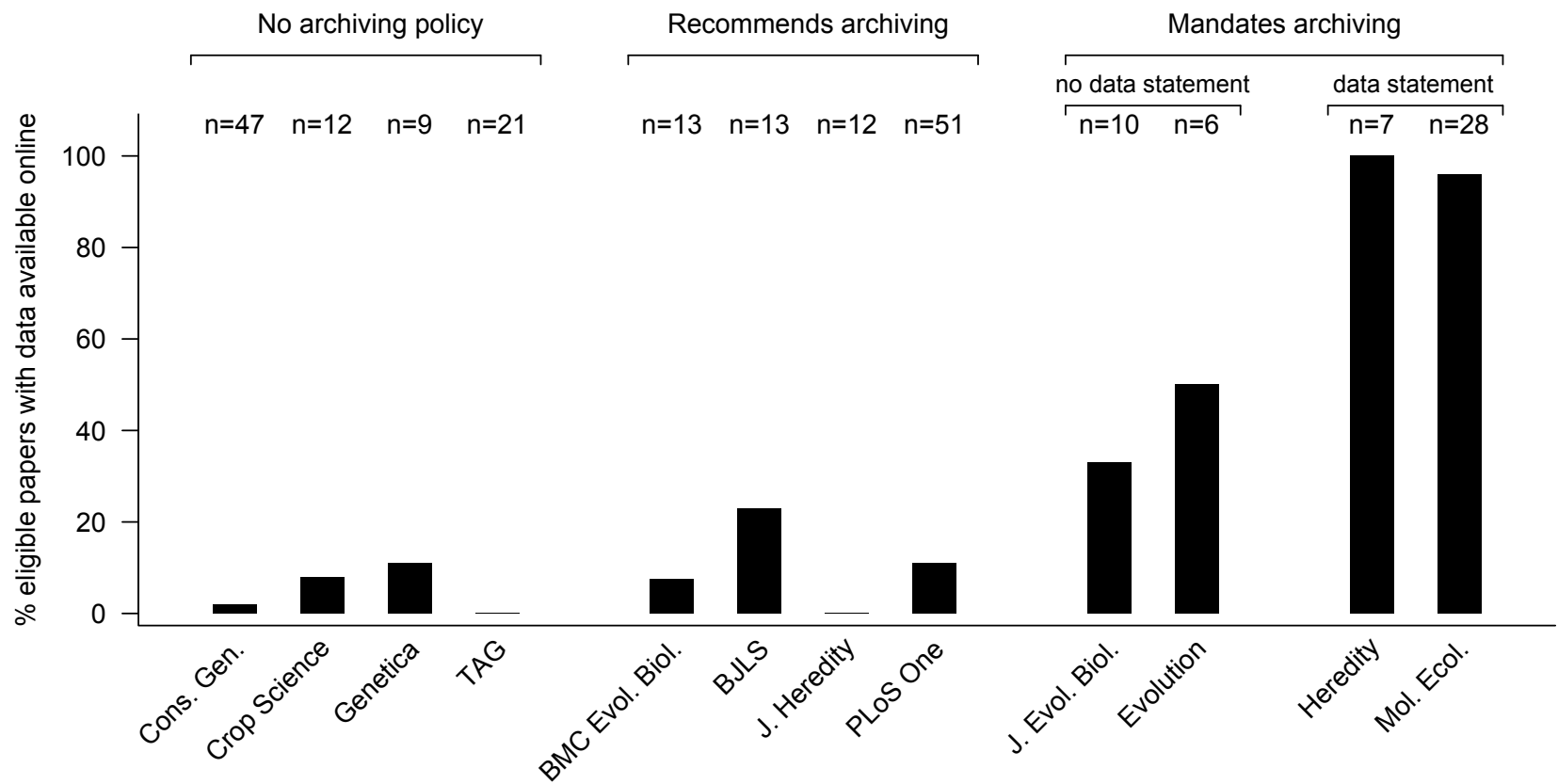

327